\documentclass[10pt,a4paper]{article}
\usepackage[tmargin=2cm,bmargin=2cm,lmargin=1.5cm,rmargin=1.5cm]{geometry}
\usepackage{amsmath}
\usepackage{multirow}
\usepackage{amssymb}
\usepackage{graphicx}
\usepackage{cite}
\usepackage{rotating}

\title{Magnetic interaction and anisotropy axes arrangement in nanoparticle aggregates can enhance or reduce the effective magnetic anisotropy}

\author{V. R. R. Aquino$^{1}$, L. C. Figueiredo$^{2}$, J. A. H. Coaquira$^{2}$, M. H. Sousa$^{3}$, A. F. Bakuzis$^{1,}$*}

\date{\footnotesize{$^{1}$Instituto de F\'{i}sica, Universidade Federal de Goiás, 74690-900, Goiânia-GO, Brazil\\
      $^{2}$Instituto de F\'{i}sica, Nucleo de F\'{i}sica Aplicada, Universidade de Bras\'{i}lia, 70910-900, Bras\'{i}lia-DF, Brazil\\
      $^{3}$Faculdade de Ceilândia, Universidade de Bras\'{i}lia, 72220-140,Bras\'{i}lia-DF, Brazil\\
      *Corresponding author: bakuzis@ufg.br}}

\begin{document}

\maketitle

\begin{abstract}
The magnetic response of nanostructures plays an
important role on biomedical applications being strongly influenced
by the magnetic anisotropy. In this work we investigate the role of
temperature, particle concentration and nanoparticle arrangement forming
aggregates in the effective magnetic anisotropy of Mn-Zn ferrite-based
nanoparticles. Electron magnetic resonance and coercivity temperature
dependence analyses, were critically compared for the estimation of
the anisotropy. We found that the temperature dependence of the anisotropy
follows the Callen-Callen model, while the symmetry depends on the
particle concentration. At low concentration one observes only an
uniaxial term, while increasing a cubic contribution has to be added.
The effective anisotropy was found to increase the higher the particle
concentration on magnetic colloids, as long as the easy axis was at
the same direction of the nanoparticle chain. Increasing even further
the concentration up to a highly packed condition (powder sample)
one observes a decrease of the anisotropy, that was attributed to
the random anisotropy axes configuration.
\end{abstract}

\section{INTRODUCTION}

The magnetic anisotropy constant have a great impact
in the magnetic response of nanoparticles. In non-interacting systems
it defines (together with the particle size) if the nanoparticle is
at the blocked or superparamagnetic (SP) regime. SP particles are
believed to play an important role in biomedical applications, spanning
from contrast agents for MRI, heat generators in magnetic hyperthermia,
cell sorting applications due to magnetophoresis properties, among
others.\cite{Javed2017,STEPHEN2011,Teresa2019,RodriguesIJH13}
Moreover nanoparticles can form aggregates, for example linv ear chains
or spherical nanostructures (even at the SP state due to van der Waals
interactions),\cite{BakuzisACIS13}
where in this case the intraparticle interactions (mainly due to magnetic
dipolar interactions) can modify the effective magnetic anisotropy.
The knowledge of this effect is of great fundamental and technological
importance due to the several applications of magnetic nanoparticles
(even beyond the biomedical field), since it directly correlates with
the relaxation of the magnetization.

In the literature there is still a strong debate
about the effect of dipolar interaction, where some authors claim
it decreases the effective anisotropy, while others point to the opposite
effect.\cite{BakuzisACIS13,Morup1994,BodkerPRL94,Dormann1988,Dormann1996,BranquinhoSR13}
It is also curious to notice that few articles discuss about the task
of anisotropy axes arrangement, the anisotropy temperature dependence
or the particle interaction role on the anisotropy energy symmetry,
which again can impact the magnetic relaxation.\cite{CoffeyJAP12}

Nowadays several techniques are applied to determine
the magnetic anisotropy, as for instance ZFC/FC magnetization curves,\cite{knobelASP,Zeb2018}
coercivity temperature dependence analysis,\cite{DenardinPRB,Mendon2012}
electron magnetic resonance,\cite{BakuzisJMR96,BakuzisJAP99,GAZEAU1998175}
Mossbauer spectroscopy,\cite{Morup1994}
among others. In some cases one first determine the blocking temperature
and then calculate the magnetic anisotropy, where some methods assume
the anisotropy to be temperature independent. This is obvious incorrect
since several anisotropy contributions are temperature dependent,
such as the magnetostatic or magnetoelastic terms. In particular,
for the \emph{ZFC/FC} curves,\cite{knobelASP}
generally, it is not clear that the determination of anisotropy via
blocking temperature is an estimated anisotropy refering to the value
at this temperature. For several studies this might not be that important,
however there is a great interest on using magnetic nanoparticles
for cancer therapy through magnetic hyperthermia.\cite{Teresa2019,Jordan1997,Rodrigues2017,RodriguesIJH13}
Here temperature effects can have a great impact on the clinical outcome.
Furthermore this biomedical application showed a strong dependence
upon particle arrangement,\cite{BranquinhoSR13,DennisAdvFunMat15,Hogan2016,Bischof2014,Dicoroto2014,Roubeau2015,Edouard2011}
and therefore understanding the role of particle interactions and
nanoparticle arrangements is also crucial. It might also be relevant
to notice the experimental condition that the effective anisotropy
is determined, for example in some works the magnetic anisotropy of
the nanoparticles is estimated using a powder sample, while the relevant
application property is analysed for the colloidal suspension. In
the powder sample the nanoparticles are highly packed and one might
assume that the anisotropy axes are randomly arranged, while in the
colloidal suspension distinct aggregate formations can arise, as for
instance linear chains that are expected to have their anisotropy
axes arranged along the chain (longitudinal configuration).\cite{Carria2017}

In this work, we investigate the role of particle
arrangement on the effective magnetic anisotropy by critically comparing
data using the coercivity temperature dependence analysis and electron
magnetic resonance (EMR). In the $Hc$ vs $T$ method powder samples
were analysed, while in the former magnetic colloids at distinct particle
concentrations were investigated. Mn-Zn ferrite nanoparticles surface-coated
with citric acid of distinct sizes were compared, allowing us to determine
the value of the effective magnetic anisotropy from the non-interacting
condition (highly diluted magnetic colloid) up to a particle volume
fraction of around 0.64, that corresponds to the packing fraction
of monodisperse spherical particles.\cite{Donev2004}
Here, we demonstrate that magnetic anisotropy is strongly temperature
dependent, and that its behavior is well represented by the Callen-Callen
model.\cite{CallenJPCS}
The room temperature anisotropy is found to increase the higher the
particle concentration in the colloid, and above a critical concentration
it shows a cubic anisotropy symmetry contribution that was not reported
before in other works. We show that the experimental result is in
accordance with the theoretical prediction that the existence of linear
chains will influence the anisotropy by means of an additional uniaxial
contribution.\cite{BakuzisACIS13}
On the other hand, the existence of the cubic anisotropy term suggests
a multipolar contribution. Increasing even further the concentration
up to a highly packed condition (powder sample), we observed a decrease
of the anisotropy and relate it to the random anisotropy axes configuration
for this particle arrangement situation. The result might be useful
on understanding some contradictory reports in the literature regarding
the effect of particle interactions, and might impact not only the
magnetic hyperthermia field, but also magnetic particle imaging\cite{GleichNature05,GoodwillAdvMat12}
and magnetic nanothermometry,\cite{WeaverMedPhys09,ZhongSR14}
among others.

The article is organized as follows: In section
II, we present the theoretical background for both
methods, coercivity temperature dependence analysis and electron magnetic
resonance, used to determine the magnetic anisotropy. Different from
most works of the literature the effect of the temperature dependence
of the anisotropy is explicitly taken into account in the $Hc$ vs
$T$ method, that also determines the sample blocking temperature
distribution. On the other hand, EMR extracts the anisotropy field.
Where from the temperature dependence magnetization studies of the
samples one can determine the effective magnetic anisotropy as function
of temperature and particle concentration. Section III
discuss the experimental procedures, i.e. synthesis and characterization
techniques. Finally, section IV
presents the results and discussions, while the conclusions are shown
in section V.

\section{THEORETICAL BACKGROUND}

\subsection{Coercivity temperature dependence method}

According to this method, the coercivity temperature
dependence $\left\langle H_{C}\right\rangle _{T}$
of the sample can be modeled using,\cite{NunesPRB,GarciaJMMM}

\begin{equation}
\left\langle H_{C}\right\rangle _{T}=\frac{M_{R}(T)}{\chi_{sp}(T)+\frac{M_{R}(T)}{H_{CB}(T)}},\label{1}
\end{equation}
where $M_{R}=\alpha Ms\int_{T}^{\infty}P(T_{Bl})\,dT_{Bl}$ is the
remanent magnetization, $\chi_{sp}=\frac{25M_{S}^{2}}{3K_{ef}}\int_{0}^{T}T_{Bl}P(T_{Bl})\,dT_{Bl}$
the  superparamagnetic susceptibility, $P(T_{Bl})$ is the lognormal distribution function
of the blocking temperatures ($T_{Bl}$). For randomly oriented particle systems $\alpha=0.48$,
while $H_{CB}$ is described by the following equation

\begin{equation}
H_{CB}(T)=\alpha\frac{2K_{ef}(T)}{Ms}\left(1-\left(\frac{T}{\left\langle T_{Bl}\right\rangle _{T}}\right)^{3/4}\right),\label{2}
\end{equation}
with $\left\langle T_{Bl}\right\rangle _{T}=\frac{\int_{T}^{\infty}T_{Bl}P(T_{Bl})\,dT_{Bl}}{\int_{T}^{\infty}P(T_{Bl})\,dT_{Bl}}$ a temperature dependent parameter related to the
distribution of the blocking temperatures at a given temperature $T$,
while the exponent $3/4$ corresponds to the random anisotropy case.\cite{GarciaJMMM}
More important, different from previous works in the literature, the
effective anisotropy constant is assumed to be temperature dependent.
This dependence is shown to experimentally follow the Callen-Callen model,\cite{CallenJPCS}
which establishes a relationship between magnetization and anisotropy
via the equation:

\begin{equation}
\frac{K(T)}{K(0)}=\left[\frac{Ms(T)}{Ms(0)}\right]^{l(l+2)/2},\label{3}
\end{equation}
For the uniaxial case $l=2$ and $K(0)$ is the anisotropy at $T=0\,K$.
Thus, this equation can be rewritten using the Bloch model, $Ms(T)=Ms(0)\left(1-bT^{3/2}\right)$,
where $Ms(0)$ is the saturation magnetization at $0\,K$ and $b$
is the Bloch constant, that are
determined from magnetization measurements. Therefore, the temperature
dependence of the effective uniaxial magnetic anisotropy constant
can be written as
\begin{equation}
K_{ef}(T)=K(0)\left(1-bT^{3/2}\right)^{3}.\label{4}
\end{equation}

\subsection{EMR method}

The other technique that will be used to estimate
the magnetic anisotropy is the Electron Magnetic Resonance (EMR).
For spherical particles the resonance field condition is given by \cite{BakuzisJMR96,BakuzisJAP99}

\begin{equation}
H_{R}=\frac{\omega}{\gamma\sqrt{1-\alpha^{2}}}-\frac{2K_{ef}}{Ms},\label{5}
\end{equation}
where $\omega$ is the angular frequency, $\gamma$ is the gyromagnetic
ratio and $\alpha$ is the damping
term (usually much lower than 1). Note that the effective anisotropy
constant can be expanded in terms of spherical harmonics such as $K_{ef}=\sum_{l}\sum_{m}K_{l}P_{l}(cos\theta)e^{im\varphi}$,
where $P_{l}$ is the Legendre
polynomial and $\theta$ is the angle between the applied magnetic
field and the nanoparticle anisotropy axis.\cite{BakuzisJMR96}
For spherical particles, $m=0$, while considering
only the uniaxial case, $l=2$ term, reveals an uniaxial anisotropy
field $2K_{2}^{s}/M_{S}P_{2}(cos\theta)$. For
the longitudinal case ($\theta=0)$ this expression is the well known
uniaxial anisotropy field $2K_{2}^{s}/M_{S}$, as
expected. On the other hand, more complex cases can appear by including
other values of $l$, that reflect the
symmetry of the anisotropy. For example, the additional existence
of a cubic contribution, $l=4$, results in the condition: 

\begin{equation}
H_{R}=\frac{\omega}{\gamma\sqrt{1-\alpha^{2}}}-\frac{2K_{2}}{Ms}P_{2}(cos\theta)-\frac{2K_{4}}{Ms}P_{4}(cos\theta).\label{6}
\end{equation}

So, the behavior of $Hr$ as a function of the angle
can indicate the anisotropy symmetry, where including only the $l=2$
term, reflects an uniaxial anisotropy symmetry, while the necessity
of other contributions reveals a more complex situation.

\subsection{Dipolar interaction contribution to the anisotropy}

In 2013, Bakuzis et al.\cite{BakuzisACIS13}
mathematically showed that the dipolar interaction between nanoparticles,
forming small agglomerates in a linear chain, has a uniaxial contribution
due to the particle-particle interaction term. The
linear chain model was applied to two situations, namely fanning and
coherent. In the coherent case, it is assumed that the magnetic moments
of the particles are all in the same direction, rotating coherently
in the direction of the magnetic field. In the fanning structure,
the magnetic moments of the adjacent particles rotate in opposite
directions. In both cases, fanning and coherent, a uniaxial contribution
to the energy density of the particle ($l=2$ term)
is noted.\cite{BakuzisACIS13}
These results can be generalized for the case of a chain containing
$Q$ particles. Below, we present the contribution to effective anisotropy
constant in the fanning case,\cite{BakuzisACIS13,BranquinhoSR13}

\begin{equation}
K_{dip}^{fanning} = \dfrac{\mu_{0}}{4\pi}\frac{M_{S}^{2}V_{p}}{\left(\bar{D}+d_{ss}\right)^{3}}\nonumber
 \times \left(\sum_{i=odd}^{Q}\frac{\left(Q-i\right)}{Q\left(i\right)^{3}}+3\sum_{i=pair}^{Q}\frac{\left(Q-i\right)}{Q\left(i\right)^{3}}\right),\label{7}
\end{equation}
where $\mu_{0}$ is a magnetic permeability of the vacuum,$\overline{D}$
is the nanoparticles mean diameter and $V_{P}$ the particle volume.
The nanoparticles centers are distant from a value $r=\bar{D}+d_{ss}$,
where $d_{ss}$ is the distance between the surfaces of the nanoparticles. The coherent calculation can be found in.\cite{BakuzisACIS13}

\section{EXPERIMENTAL PROCEDURE}

Among the various methods for synthesizing magnetic
nanoparticles, with new morphologies and dimensions, the hydrothermal
technique has been extensively explored. This synthetic route allows
working with temperatures above the boiling point of the chosen solvent,
changing the crystallization/recrystallization conditions in the synthetic
environment.\cite{Byrappa2013} For
the preparation of magnetic nanoparticles based on manganese-zinc
ferrite with the stoichiometry $\mathrm{Mn_{0.75}Zn_{0.25}Fe_{2}O_{4}}$,
we mixed $3.75\:mmol$ of manganese chloride tetrahydrate $(\mathrm{MnCl_{2}4H_{2}O})$,
$1.25\:mmol$ of ions zinc chloride ($\mathrm{ZnCl_{2}}$) and $10\:mmol$
of ferric chloride hexahydrate $(\mathrm{FeCl_{3}6H_{2}O})$ from
$1\:mol/l$ stock solutions. Thus, $50\:ml$ of $8.0\mathrm{\,wt\%}$
aqueous methylamine $\mathrm{(CH_{3}NH_{2}})$ was poured into the
metal's solution under magnetic strring for about $5\:min$ at room
temperature. The mixture was sealed in a teflon-lined autoclave and
maintained at $160\:^\circ \text{C}$ for $5\:h$ inside an oven.

After this time, the formed magnetic material was
washed three times with distilled water. To the resulting magnetic
material was added $300\:ml$ To prepare the citrate-capped nanoparticles,
$2\:g$ of trisodium citrate was added to this dispersion under magnetic
stirring at $80\,{^\circ}C$ for $30\:minutes$. The precipitate was
magnetically separated, washed with acetone and redispersed in water.
The pH of dispersion was carefully adjusted to form a stable magnetic
fluid at pH $\sim7$. To obtain samples with different
mean diameters, we used a size-sorting method based on the increase
the ionic strength of the sol -- through the addition of NaCl --
which induces a phase transition in the colloid.\cite{SousaMicrochem11,MassartJMMM}
Typically, NaCl is added to the sol with a magnet placed at the bottom of the
flask until visual colloidal separation. After magnetic separation,
the supernatant (with smallest nanoparticles) and precipitate (with
larger nanoparticles) are washed with acetone in order to resuspend
nanoparticles in aqueous solution at pH $\sim7$.

Once the magnetic fluid was achieved, they were diluted in different
volumes fractions, which were checked by expression: $\phi\approx M_{fluid}/Ms$,
where$M_{fluid}$ and $Ms$ are the saturation magnetization of the
fluid and the powder, using a VSM (vibrating sample magnetometer,
$2\,Tesla$). After obtaining the magnetic fluid, 
the nanoparticles were characterized by several techniques, such as
energy dispersive spectroscopy (EDS), obtaining images for compositional
analysis; X-ray diffraction (XRD), to obtain the crystalline phase
and the average size of the crystallite; transmission electron microscopy
(TEM), for the calculation of the distribution of diameters and shape
of the particles; Superconducting Quantum Interference Device (SQUID)
magnetometer, necessary to achieve saturation magnetization and the
coercive field at low temperatures and electron magnetic resonance
(EMR), for anisotropy field determination.

Measurements of the chemical composition of the
samples were carried out using energy dispersive spectroscopy (EDS)
using the transmission electron microscope (JEOL model JEM-2100),
operating in EDS mode at $15\:kV$. For the determination
of particle sizes, we dry a part of the colloid to obtain the powder
and perform the analysis by XRD (Shimadzu 6000). Images of the nanoparticles
were obtained using the transmission electron microscope (Jeol model
JEM-2100) operated at $200\:kV$, with resolution of $25\:\mathring{\mathrm{A}}$.
For the characterization of magnetic properties at low temperatures,
we used a VSM-SQUID (Quantum Design PPMS3) with a DC field ranging
from -70 to 70 $kOe$, and temperatures ranging from 5 to 300 K. On
the other hand the calculation of the particle volume fraction at
room temperature was obtained using a VSM (ADE Magnetic, EV-9 model)
with a DC field ranging from -20 to 20 $kOe$. Finally, EMR measurements
were perfomed with a spectrometer EMX-Plus model Bruker, where the
magnet had a magnetic field amplitude up to $14\:kG$ and X-band microwave
bridge tuned around 9.5GHz. The EMR procedure to extract the anisotropy
field was the same of Refs.\cite{BakuzisJMR96,BakuzisJAP99,BakuzisJMMM01}.
Basically, at room temperature, one apply the highest external magnetic
field to the magnetic fluid sample with the objective to orient the
nanoparticle magnetic anisotropy axis along the field direction. With
the field on, the sample is frozen to 100K. The procedure blocks the
nanoparticle's anisotropy in a specific direction. A goniometer device
allow rotation of the sample with respect to the applied field during
the EMR experiment. So, EMR spectra at distinct angles, for a given
temperature and particle concentration, can be obtained to determine
the value and symmetry of the effective anisotropy constant.

\section{RESULTS AND DISCUSSION}

Figure 1(a) shows the XRD data
for both nanoparticles confirming the cubic spinel structure. Crystallite
sizes of 10.3 and 11.4 $nm$ were obtained using the Scherrer
equation, i.e. $D_{XRD}=\kappa\lambda/\beta cos\psi$, where $\kappa=0.89$
is the Scherrer constant, $\lambda=0.154\:nm$ is the X-ray wavelength,
$\beta$ is the line broadening in radians obtained from the square
root of the difference between the square of the experimental width
of the most intense peak to the square of silicon width (calibration
material), and $\psi$ is the Bragg angle of the most intense peak.
The inset of Fig. 1(b) shows
an image of a film made of 10 nm nanoparticles, while some spots show
places where were performed the EDS-TEM analysis. Fig. 1b shows the
EDS analysis of this sample, where one can clearly observe the existence
of Mn, Zn and Fe, as expected (Cu signal is due to the TEM microgrids).
Fig 1(c) shows the size distribution
for both samples obtained from the analysis of TEM pictures, while
the inset shows TEM images of the 10nm size nanoparticles revealing
spherical-like particles. From the fit of the histogram using the
lognormal distribution size (median and size dispersion parameters)
one can calculate the mean diameters and the standard deviation, $10\:\pm\:2$
and $11\:\pm\:2$ $nm$. Magnetic characterization is shown in Figure
1(d) for the 10nm particle size,
while the inset shows hysteresis curves at low field range. Measurements
were performed at a wide temperature range 5 to 300 K for a powder
sample. The saturation magnetization value is obtained from the analysis
at the high field limit, i.e. from extrapolation of data of $M\,\times\,1/H$
when $1/H$ tends to zero. Fig. 1(e)
shows the temperature dependence of the saturation magnetization for
both samples. The symbols represent the experimental data, while line
is the best fit using the Bloch's law, that revealed the Bloch constants,
$7.5\,\times\,10^{-5}$ and $6.6\,\times\,10^{-5}K^{-3/2}$ for
$10$ and $11\,nm$ diameters and the saturation magnetization value
at $0\,K$, 585 and 565 $emu/cm^{3}$, respectively. On the other
hand, the particle concentration of the magnetic fluids was obtained
by the analysis of the magnetization at room temperature using a 2T
VSM. Figure 1(f) shows the magnetization
curves of the same sample but now at different particle volume fractions
($\phi$). The estimation of $\phi$ arises from the ratio of the
saturation magnetization of the sample to the saturation measured
for the nanoparticle (powder sample) at the same experimental condition,
that for our samples were found to be 293 $emu/cm^{3}$ and 303 $emu/cm^{3}$
for particles of 10 and 11 $nm$.

\begin{figure*}
\includegraphics[scale=0.8]{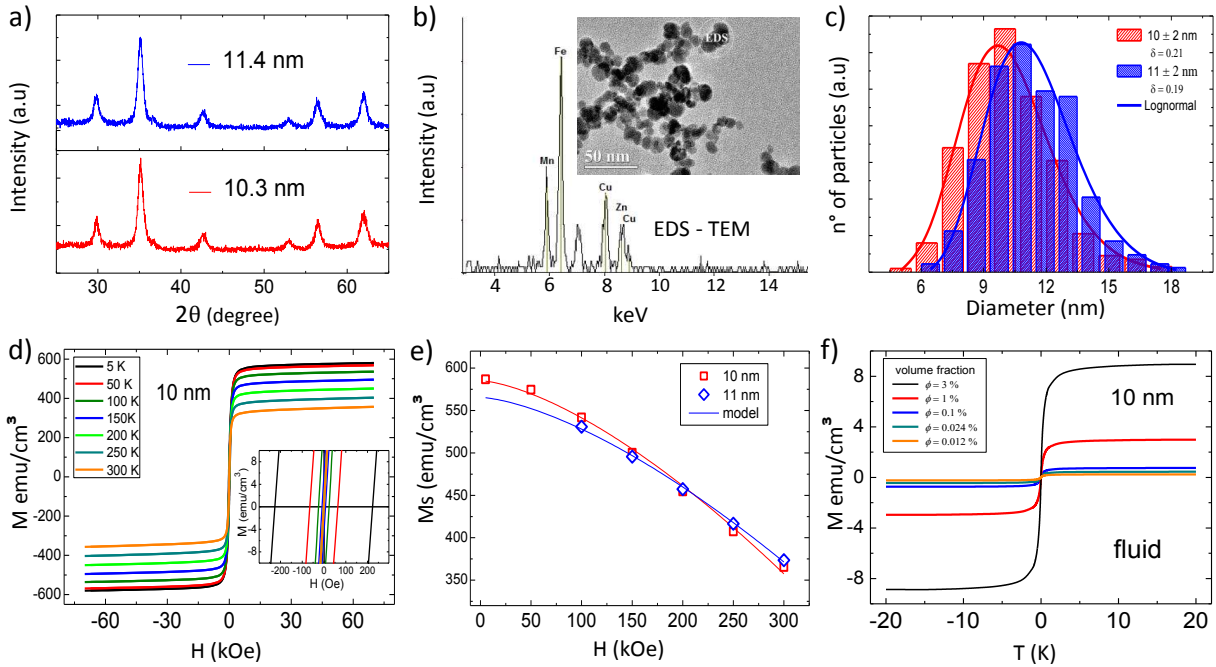}

\caption{(a) XRD data of the MnZn samples. Crystaline sizes
are determined using Scherrer equation. (b) EDS analysis of the 10nm
sample showing the existence of Mn, Zn and Fe in the nanoparticle
composition. The inset shows a TEM picture and the position of the
EDS analysis. (c) Size distribution obtained by TEM. (d) Hysteresis
curves at distinct temperatures for the 10nm sample. The inset shows
the magnetization curves at low field range. (e) Saturation magnetization
temperature dependence study. Symbols represent data, while lines
are the Bloch law model. (f) Magnetization data of the 10nm-based
colloid for particle volume fraction determination.}
\end{figure*}

\subsection{Blocking temperature distribution}

Figure 2(a) shows the temperature dependence of the coercivity for both samples,
while in the inset is shown the low field hysteresis curves for the
11nm sample. Symbols correspond to experimental data, while the lines
correspond to the best fit using the theoretical model discussed in
section IIA, that included the
effective anisotropy temperature dependence by using $K_{ef}(T)=K(0)\left(1-b\,T^{3/2}\right)^{3}$.
It is clear in this model that only the value of
the anisotropy at $0\,K$ becomes one of the parameters of adjustment
of the coercive field data, where the other fitting parameter is related
to the blocking temperature distribution, $T_{Bl}^{m}$.
Note that because of lognormal distribution properties, one can assume
that the dispersion of blocking temperatures (that is proportional
to the particle volume) is related to the size dispersity (obtained
from TEM analysis) through $\sigma_{B}=3\sigma_{TEM}$. Therefore,
the mean blocking temperature of the sample can be calculated using
the equation $\overline{T}_{Bl}=T_{Bl}^{m}\,exp[(\sigma_{B})^{2}/2]$.
Fig 2(b) shows the blocking temperature
distribution obtained from the analysis of $\left\langle Hc\right\rangle _{T}$
$vs$ $T$. The dashed lines indicate the position of the mean blocking
temperature for both samples. As expected, higher value was found
for the larger particle size. 

\begin{figure*}
\centering
\includegraphics[scale=0.8]{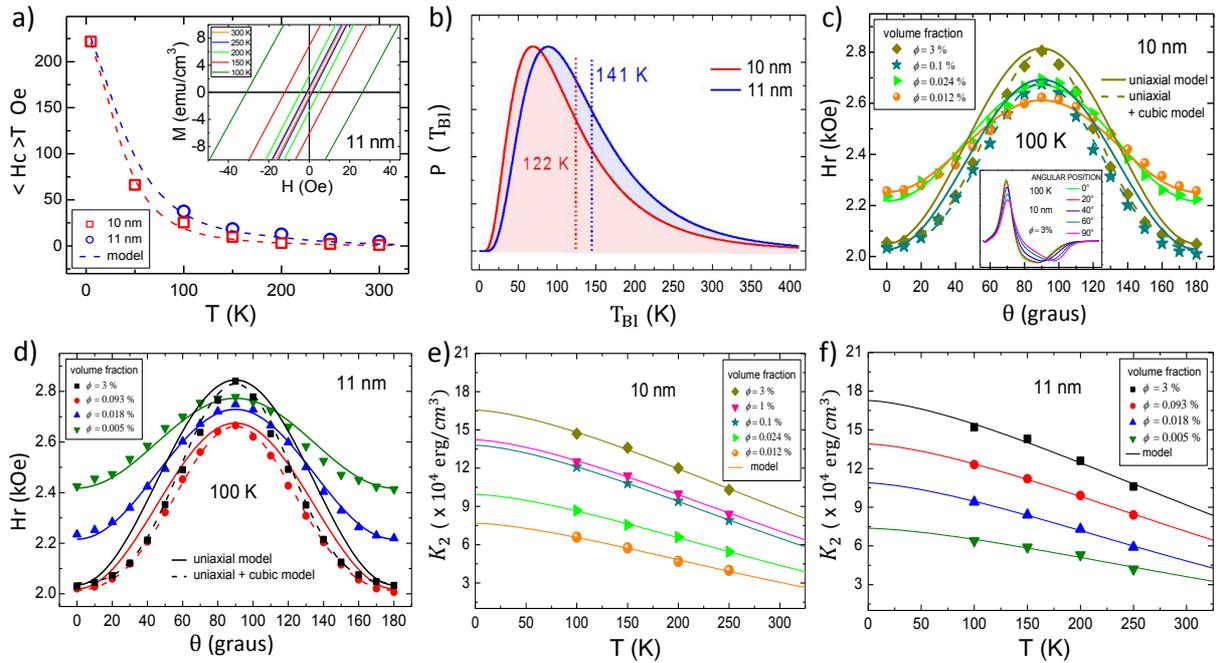}

\caption{(a) Coercive field as function of temperature. Symbols are data, while
lines correspond to the model. The inset shows the the
magnetization curves at low field range for the 11nm sample. (b)
Blocking temperature distributions obtained from the model for both
samples. In (c) and (d) EMR field as function of the angle between
the applied field and the anisotropy axis at 100K and distinct particle
concentrations, respectively for the 10nm and 11nm samples. Solid
and dash lines are the adjustments of the experimental data according
to the models discussed in the text. The inset in (c)
shows typical EMR spectra at distinct angles. In (e) and (f) symbols
represent the uniaxial anisotropy term $K_{2}$ as function of temperature
and particle concentration, respectively for 10 and 11nm samples,
while lines are the best fit using the Callen-Callen model. }
\end{figure*}

Table 1 summarizes
the parameters obtained from the coercive field analysis, $\left\langle Hc\right\rangle _{T}$
$vs$ $T$ method, namely $\overline{T}_{Bl}$ and $K_{ef}(0\,K)$.
For comparison we also included the values estimated using the non-temperature
dependent model, $K_{ef}^{n}$ and $\overline{T}_{Bl}^{n}$ . Curiously
the value obtained is very close to $K_{ef}(0\,K)$, while $\overline{T}_{Bl}^{n}$
is slightly lower than $\overline{T}_{Bl}$. As for instance, the
10nm sample showed $\overline{T}_{Bl}=122K$, while for the non-temperature
dependent model $\overline{T}_{Bl}^{n}=110K$. The difference with
other models from the literature\cite{NunesPRB}
are that: (i) firstly, in our case, we can easily identify the temperature
correspondence of this anisotropy; (ii) Second, here there is no necessity
to fit ZFC/FC curves to extract the blocking temperature distribution,
a procedure that depending on the nanoparticle is not easily performed;
(iii) In the present model, one can estimate the anisotropy constant
at room temperature, where it is found that both samples shows room
temperature anisotropy values on the order of $\sim10^{4}~erg/cm^{3}$.
However, it is very important to notice that so far we made the asumption
that the anisotropy temperature dependence follows the Callen-Callen
model. In the next section we demonstrate that this is indeed a very
good approximation.

\begin{table}
\caption{Effective anisotropy constants and average blocking temperatures of
the $\mathrm{Mn_{0.75}Zn_{0.25}Fe_{2}O_{4}}$samples according to
the coercivity temperature dependence analysis.}
\centering
\begin{tabular}{cccccc}
 
\begin{tabular}{c}
{$\overline{D}$}\tabularnewline
{$(nm)$}\tabularnewline
\end{tabular} & %
\begin{tabular}{c}
{$K_{ef}^{n}$}\tabularnewline
{$(erg/cm^{3})$}\tabularnewline
\end{tabular} & %
\begin{tabular}{c}
{$\overline{T}_{Bl}^{n}$}\tabularnewline
{$(K)$}\tabularnewline
\end{tabular} & %
\begin{tabular}{c}
{$K_{ef}(0\,K)$}\tabularnewline
{$(erg/cm^{3})$}\tabularnewline
\end{tabular} & %
\begin{tabular}{c}
{$K_{ef}(300\,K)$}\tabularnewline
{$(erg/cm^{3})$}\tabularnewline
\end{tabular} & %
\begin{tabular}{c}
{$\overline{T}_{Bl}$}\tabularnewline
{$(K)$}\tabularnewline
\end{tabular}\tabularnewline
\hline &  &  &  &  & \tabularnewline

\begin{tabular}{c}
{\small 10}\tabularnewline
\end{tabular} & 
\begin{tabular}{c}
{\small $1.5\times10^{5}$}\tabularnewline
\end{tabular} & 
\begin{tabular}{c}
{\small $110\,\pm\,76$}\tabularnewline
\end{tabular} & 
\begin{tabular}{c}
{\small $1.5\times10^{5}$}\tabularnewline
\end{tabular} & 
\begin{tabular}{c}
{\small $3.4\times10^{5}$}\tabularnewline
\end{tabular} & 
\begin{tabular}{c}
{\small $122\,\pm\,85$}\tabularnewline
\end{tabular}\tabularnewline

\begin{tabular}{c}
\tabularnewline
{\small 11}\tabularnewline
\end{tabular} & 
\begin{tabular}{c}
\tabularnewline
{\small $1.4\times10^{5}$}\tabularnewline
\end{tabular} & 
\begin{tabular}{c}
\tabularnewline
{\small $130\,\pm\,80$}\tabularnewline
\end{tabular} & 
\begin{tabular}{c}
\tabularnewline
{\small $1.4\times10^{5}$}\tabularnewline
\end{tabular} & 
\begin{tabular}{c}
\tabularnewline
{\small $4.1\times10^{4}$}\tabularnewline
\end{tabular} & 
\begin{tabular}{c}
\tabularnewline
{\small $141\,\pm\,87$}\tabularnewline
\end{tabular}\tabularnewline
\end{tabular}
\end{table}

\subsection{Anisotropy temperature dependence and the Callen-Callen model}

The inset of Fig. 2(c)
shows EMR spectrum at different angle positions for the 10nm sample
with a particle volume fraction of 3\%. Similar experiments were performed
at distinct particle concentrations for both samples. Fig. 2(c)
shows the resonance field position as function of the angle between
the applied field and the anisotropy axis for the 10nm sample at different
particle concentrations, while Fig. 2(d)
shows similar data but for the 11nm sample. It is clear that there
is a shift in the resonance field position increasing the angle up
to 90 degrees. Above this value the resonance field position decreases
returning to its initial position for an angle of 180 degrees. The
EMR measurements shown here were performed at 100 K, but other experiments
at temperatures 150, 200 and 250K were also obtained. Higher temperatures
were not analysed because the nanoparticles were dispersed in water,
and we wanted to mantain the nanoparticles in the frozen matrix and
avoid the solid-liquid transition. Symbols represent EMR data, while
solid lines corresponds to the uniaxial case (only $l=2$ term - $K_{2}^{s}$)
and dashed lines are the multiaxial case (Eq.(6)),
where one can obtain the values of $K_{2}$ and $K_{4}$. Note that,
for low concentrations, the uniaxial contribution alone ($K_{2}^{s}$)
is able to explain the behavior of the resonance field. In particular,
it is further noted that the increase in the difference of the Hr
variation is a result of a higher anisotropy when the concentration
increases. However, in the higher concentrations the adjustment of
the experimental data considering only the $l=2$ term was not so
good. Therefore the analysis presented here corresponds to the one
using Eq. (6), i.e. one obtain
both $K_{2}$ and $K_{4}$. According to the theoretical models the
difference between the resonance field at 0 and 90 degrees is related
to the anisotropy field. Since the saturation magnetization temperature
dependence was determined before (see Fig. 1(e)),
using the anisotropy field we extract the temperature dependence of
the anisotropy constants. Figs. 2(e)
and 2(f) shows the effective
anisotropy temperature dependence for the 10 and 11 nm samples, respectively.
Here we are showing only the $K_{2}$ value, but Table 2
summarizes all the other parameters analysed in this study (including
the $K_{2}^{s}$ for the uniaxial case). Symbols represent experimental
data, while solid lines correspond to the best fit using the Callen-Callen
model. It is obvious from this analysis the excellent aggreement with
the data, justifying the assumption on the later section. On the other
hand, $K_{4}$ is only relevant at high particle concentrations, although
its value is around one order of magnitude lower than $K_{2}$ (see
Table 2). Room temperature anisotropy
values can be found using the Callen-Callen model, and revealed an
increase the higher the particle concentration. Above a critical concentration
a cubic anisotropy symmetry contribution has to be added in the analysis,
suggesting a possible multipolar contribution to the anisotropy. Possibly,
this is the first experimental evidence of the existence of a multipolar
contribution at high concentrations. It is possible that such term
is the result of a more complex organization of nanoparticles than
only isolated linear chains. In favor of this argument is the fact,
well known in the literature, that the magnetic fluid has a liquid-solid
transition increasing the concentration of particles.\cite{EloiPRE10,KlokkenburPRL}
This phenomenon results in the formation of complex self-organized
structures, for example in the formation of hexagonal columnar structures.\cite{KlokkenburPRL}
This might be different from the case of isolated linear chains, that
due to the dipolar interaction between nanoparticles showed only a
uniaxial contribution term.\cite{BakuzisACIS13}

\begin{sidewaystable}
\centering
\caption{Anisotropy constants of $\mathrm{Mn_{0.75}Zn_{0.25}Fe_{2}O_{4}}$
of 10 and 11$nm$ for each concentration and at different temperatures.
Thus, from 100 to 250 $K$ EMR data, while 0 and 300 $K$ corresponds
to extrapolation values using Callen-Callen model.}

\vspace{1.0cm}
\begin{tabular}{cccccccc}
\begin{tabular}{c}
 $\phi$\tabularnewline
\tabularnewline
 $(\%)$\tabularnewline
\end{tabular} & 
\begin{tabular}{c}
 $\overline{D}$\tabularnewline
\tabularnewline
 $(nm)$\tabularnewline
\end{tabular} & 
\begin{tabular}{c||c||c}
\multicolumn{3}{c}{ $0\,K$}\tabularnewline
\hline 
\multicolumn{3}{c}{$K_{2}$}\tabularnewline
\multicolumn{3}{c}{$(\times10^{4}erg/cm^{3})$}\tabularnewline
\end{tabular} & 
\begin{tabular}{ccc}
\multicolumn{3}{c}{ $100\,K$}\tabularnewline
\hline 
 $K_{2}$ &  $-K_{4}$ &  $K_{2}^{s}$\tabularnewline
\multicolumn{3}{c}{ $(\times10^{4}erg/cm^{3})$}\tabularnewline
\end{tabular} & 
\begin{tabular}{ccc}
\multicolumn{3}{c}{ $150\,K$}\tabularnewline
\hline 
 $K_{2}$ &  $-K_{4}$ &  $K_{2}^{s}$\tabularnewline
\multicolumn{3}{c}{ $(\times10^{4}erg/cm^{3})$}\tabularnewline
\end{tabular} & 
\begin{tabular}{ccc}
\multicolumn{3}{c}{ $200\,K$}\tabularnewline
\hline 
 $K_{2}$ &  $-K_{4}$ &  $K_{2}^{s}$\tabularnewline
\multicolumn{3}{c}{ $(\times10^{4}erg/cm^{3})$}\tabularnewline
\end{tabular} & 
\begin{tabular}{ccc}
\multicolumn{3}{c}{ $250\,K$}\tabularnewline
\hline 
 $K_{2}$ &  $-K_{4}$ &  $K_{2}^{s}$\tabularnewline
\multicolumn{3}{c}{ $(\times10^{4}erg/cm^{3})$}\tabularnewline
\end{tabular} & %
\begin{tabular}{ccc}
 & \multicolumn{2}{c}{ $300\,K$}\tabularnewline
\cline{2-3} 
 &  $K_{2}$ &  -$K_{4}$\tabularnewline
 & \multicolumn{2}{c}{ $(\times10^{4}erg/cm^{3})$}\tabularnewline
\end{tabular}\tabularnewline
\hline &  &  &  &  &  &  & \tabularnewline
 $3$ &  $10$ &  $16.5$ &  $14.7\quad3.3\quad13.8$ &  $13.6\quad3.0\quad12.8$ &  $12.0\quad2.6\quad11.2$ &  $10.3\quad2.3\quad9.6$ &  $8.8\quad1.70$\tabularnewline
 $1$ &  $10$ &  $14.2$ &  $12.5\quad2.2\quad11.3$ &  $11.4\quad2.3\quad10.5$ &  $10.0\quad1.8\quad9.1$ &  $8.4\quad2.3\quad7.9$ &  $7.1\quad1.19$\tabularnewline
 $0.1$ &  $10$ &  $13.8$ &  $12.1\quad2.1\quad10.6$ &  $10.8\quad1.8\quad9.8$ &  $9.4\quad1.3\quad8.7$ &  $7.9\quad1.1\quad7.6$ &  $6.5\quad0.98$\tabularnewline
 $0.024$ &  $10$ &  $9.9$ &  $8.7\quad1.0\quad8.4$ &  $7.6\quad0.8\quad7.1$ &  $6.6\quad0.8\quad6.4$ &  $5.4\quad0.6\quad5.4$ &  $4.4\quad0.46$\tabularnewline
 $0.012$ &  $10$ &  $7.8$ &  $6.6\quad0.3\quad6.5$ &  $5.7\quad0.2\quad5.6$ &  $4.7\quad0.3\quad4.8$ &  $4.0\quad0.3\quad3.9$ &  $3.2\quad0.22$\tabularnewline
 &  &  &  &  &  &  & \tabularnewline
 $3$ &  $11$ &  $17.3$ &  $15.2\quad3.6\quad14.6$ &  $14.3\quad3.3\quad13.3$ &  $12.6\quad2.7\quad12.0$ &  $10.6\quad2.2\quad10.1$ &  $9.5\,\quad1.77$\tabularnewline
 $0.093$ &  $11$ &  $13.9$ &  $12.2\quad1.9\quad14.6$ &  $11.2\quad1.7\quad10.6$ &  $9.9\quad1.4\quad9.5$ &  $8.3\quad1.0\quad8.2$ &  $7.2\,\quad0.76$\tabularnewline
 $0.018$ &  $11$ &  $10.9$ &  $9.4\quad0.8\quad9.1$ &  $8.4\quad0.6\quad7.9$ &  $7.3\quad0.4\quad7.1$ &  $5.9\quad0.2\quad5.8$ &  $4.9\,\quad0.13$\tabularnewline
 $0.005$ &  $11$ &  $7.5$ &  $6.4\quad0.3\quad6.5$ &  $5.9\quad0.2\quad5.8$ &  $5.3\quad0.1\quad5.1$ &  $4.2\quad0.1\quad4.1$ &  $3.6\,\quad0.04$\tabularnewline
\end{tabular}
\end{sidewaystable}

\subsection{The role of the magnetic interaction and axes arrangement on the anisotropy}

\begin{figure}
\centering
\includegraphics[scale=0.85]{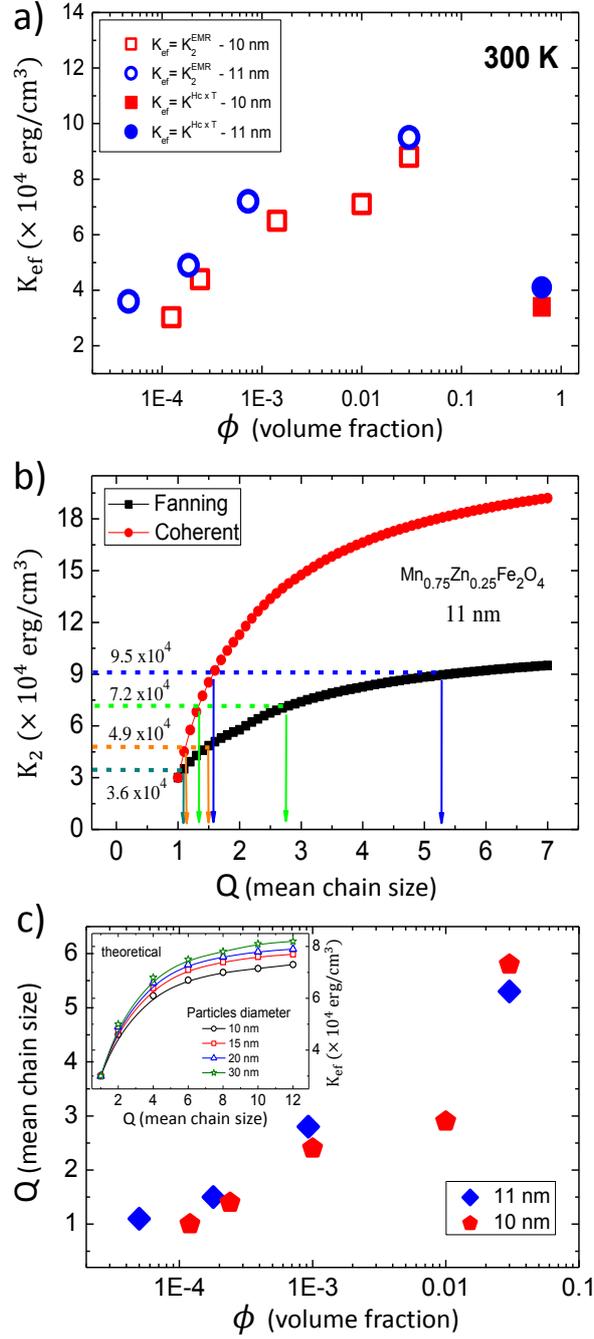}

\caption{(a) Room temperature uniaxial anisotropy constant as function of particle
volume fraction for both samples. Open symbols corresponds to dat
obtained by $EMR$ method, while solid symbols are obtained from the
$Hc\:vs\:T$ method. (b) Theoretical anisotropy constant calculation
as function of chain size. Circles corresponds to coherent, while
squares are the fanning case. Here we used $d_{ss}=1.1\,nm$ and $Ms=366\,emu/cm^{3}$.
Arrows indicate the chain size corresponding to the experimental effective
anisotropy value (dash lines). (c) Mean chain size as function of
particle volume fraction for both samples assuming the fanning case.
The inset shows the same as (b) for the fanning case, but including
distinct particle sizes.}
\end{figure}

The excellent agreement between the experimental data and the Callen-Callen
model allowed us to calculate the anisotropy constants
by estrapolation at $0\,K$ and $300\,K$ for both samples.
Table 2 summarizes all the results
obtained as function of temperature and concentration. Also, since
the main anisotropy contribution arises from the uniaxial term for
most samples, from now on we will focus on the room temperature concentration dependence of $K_{2}$. Figure
3(a), presents the data at the ambient temperature of the effective
anisotropy constant (considering only the uniaxial term - $K_{2}$),
for different samples, as a function of the particle
volume fraction. Open symbols corresponds to the analysis obtained
from the EMR data of the colloids, while solid symbols correspond
to the coercivity temperature dependence analysis (of the powder).
According to the literature,\cite{Donev2004}
for monodisperse spherical nanoparticles one can assume that the packing
of the nanoparticles in the powder configuration corresponds to a
particle volume fraction close to $\phi=0.64$ (although
the nanoparticles are not perfectly spherical or monodisperse this
is considered to be a good approximation). Recall that unlike the
case of the magnetic fluid in which the anisotropic axes are oriented
towards the freezing field (longitudinal case \cite{Carria2017}),
in the case of the powdered sample the axes are estimated to be in
the random configuration. For the colloid one can observe an increase
of the effective magnetic anisotropy the higher the particle concentration.
This result is in agreement with the theoretical prediction for the
case of nanoparticles forming a linear chain.\cite{BakuzisACIS13}

On the other hand, for the higher concentration,
a lower value is observed for the anisotropy, which at first can
suggest the existence of a maximum as a function of the concentration.
Indeed, the random anisotropy model, already applied even for magnetic
dipolar fluids,\cite{GONZAPhysRevE}
suggests a decrease of the effective anisotropy, such that for the
case of $N$ particles interacting collectively, the anisotropy can
be represented as $K_{ef}=K_{int}/\sqrt{N}$. Here $K_{int}$ refers
to the anisotropy of an individual particle (interacting or non interacting).
Therefore, the transition from a longitudinal to random condition
may explain such a decrease, at least qualitatively. As discussed
previously, to explain the existence of the cubic term in the anisotropy,
it is known in magnetic fluids that increasing the concentration of
particles occurs a liquid-solid transition.\cite{KlokkenburPRL,EloiPRE10}
In this case, formation of complex self-organized structures can break
the longitudinal condition, favoring a situation with randomly organized
anisotropy axes. The consequence of such an effect may be the decrease
in anisotropy. In the case of samples of magnetic fluids such an effect
was not observed, but may appear only in the case of higher concentrations,
which would allow the existence of a maximum. As this did not happen
for the fluids in our experimental condition, we
decided to theoretically determine, via the linear chain model,\cite{BakuzisACIS13}
if the increase in the size of the chain is able to explain the value
of the anisotropy observed experimentally.

Figure 3(b)
presents the calculations of the increase of the effective magnetic
anisotropy for a 11nm nanoparticle of $\mathrm{Mn_{0.75}Zn_{0.25}Fe_{2}O_{4}}$,
for fanning (squares) and coherent (circles) cases,
assuming a surface-to-surface distance of 1.1nm (estimated for citric
acid molecules). An increase in the effective anisotropy of the nanoparticle
is clearly evident with the increase in the number of particles in
the chain. It is also observed the effective anisotropy value tends
to saturation for a large number of particles in the chain, being
higher for the coherent case. The role of particle size is shown in
the inset of Fig. 3(c) considering
only the fanning case. From the theoretical estimations of Fig. 2(b),
by comparison with the values extracted from the EMR analysis, it
is possible to estimate the chain size for any magnetic particle concentration
of the colloids. For example, for $\phi=0.03$ we found an anisotropy
value of $K_{ef}=9.5x10^{4}erg/cm^{3}$ that for the fanning case
reveals a mean chain size of 5.3. The arrows indicate the average
chain size (x-axis) that has the experimental value of the anisotropy
(y-axis). Thus, for the volumes fractions of $0.005$, $0.018$, $0.093$
and $3$ $\%$, we obtain as average: $1.1$, $1.5$, $2.8$ and $5.3$
particles in a chain for the case fanning and $1.0$, $1.1$, $1.3$
and $1.6$ particles for the coherent case, in which it is clear the
increase of the size of the aggregate as a function of the concentration
of particles. Further smaller chains are observed for the coherent
case when compared to the fanning system. It is known, however, that
in the situation of lower energy the fanning case is more favorable.
Figure 3(c) shows the increase in the number of aggregates with the
increase of the particle volume fraction for both samples, considering
the fanning case. Similar trend is found for both
particle sizes and suggest that the increase in the effective magnetic
anisotropy is related to the formation of larger linear chains the
higher the particle concentration. This behavior is in accordance
with the theoretical model of Refs. \cite{BakuzisACIS13,BranquinhoSR13}
that states an increase of the effective anisotropy due to the dipole-dipole
particle interaction.

Finally, it is tempted to discuss what are the implications
of for cancer hyperthermia. Firstly, if the heat generation is governed
by the Néel relaxation mechanism, then is obvious that the results
presented here have important consequences, since the relaxation magnetization
depends exponentially on the effective anisotropy. So, one could in
principle tune the effective anisotropy in order to maximize the heat
generation. If one needs to increase the anisotropy then linear chains
are interesting options. Indeed, magneto-bacteria chains have been
shown to heat efficiently at high field conditions. On the other hand,
if one needs to decrease the effective anisotropy, then arranging
the nanoparticles as spherical aggregates seems an interesting approach
because the random anisotropy axes configuration could lead to this
goal. Behind this effect one can find the explanation for several
apparently contradictory results regarding the role of the magnetic
particle interaction on the heat efficiency.

\section{CONCLUSION}

Electron magnetic resonance analysis showed that
the anisotropy field of these nanoparticles at low particle concentration
show uniaxial symmetry that grows with increasing particle volume
fraction, thus confirming the influence of the dipole-dipole interaction.
At very high particle concentrations, a cubic symmetry term for the
anisotropy has to be added to the model suggesting the possibility
of multipolar contributions. It was also possible to prove that the
anisotropy of these nanoparticles is strongly temperature dependent,
and can be adjusted with the Callen-Callen model. The room-temeprature
magnetic anisotropy values obtained from EMR data analysis for the
$\mathrm{Mn_{0.75}Zn_{0.25}Fe_{2}O_{4}}$ based colloids as a function
of the concentration were the linear chain model with anisotropy axes
aligned in the chain direction, longitudinal configuration. This analysis
indicates that the increase of the anisotropy with the concentration
is related to the existence (and formation) of chains in the magnetic
colloid, and that the size of them increases with the concentration.
In particular, for the $11\,nm$ sample the anisotropy changed from
$3.6\,\times\,10^{4}~erg/cm^{3}$, corresponding to an average chain
size of only $1.1$ for the most dilute case, to an anisotropy value
of $9.5\,\times\,10^{4}$ $erg/cm^{3}$ , which in the model corresponds
to a chain containing on average $5.3$ for the most concentrated
sample ($3\,\%$ volume fraction). On the other hand,
the anisotropy of the powder sample (estimated to have a particle
volume fraction around $64\,\%$), that was evaluated from the coercive
field temperature dependence analysis, showed a reduction of the
anisotropy to $4.1\,x\,10^{4}~erg/cm^{3}$,  probably
due to the random distribution of the anisotropy axes in this experimental
condition. The results indicate that the effective magnetic anisotropy
is strongly dependent on the magnetic interaction between the particles
and the arrangement of the anisotropy axes, which might explain some
contradictory discussions in the literature since one might enhance
or decrease the effective anisotropy depending on the specific situation.\\

\section*{ACKNOWLEDGMENTS}
The authors would like to thank financial support
from the Brazilian agencies CNPq, CAPES, FAPEG, FAPDF and FUNAPE.

\bibliographystyle{unsrt}
\bibliography{References}

\begin{thebibliography}{10}

\bibitem{Javed2017}
Y.~Javed, K.~Akhtar, H.~Anwar, and Y.~Jamil.
\newblock Mri based on iron oxide nanoparticles contrast agents: effect of
  oxidation state and architecture.
\newblock {\em Journal of Nanoparticle Research}, 19:366, 2017.

\bibitem{STEPHEN2011}
Z.~R. Stephen, F.~M. Kievit, and M.~Zhang.
\newblock Magnetite nanoparticles for medical mr imaging.
\newblock {\em Materials Today}, 14:330 -- 338, 2011.

\bibitem{Teresa2019}
B.~T. Mai, P.~B. Balakrishnan, M.~J. Barthel, F.~Piccardi, D.~Niculaes,
  F.~Marinaro, S.~Fernandes, A.~Curcio, H.~Kakwere, G.~Autret, R.~Cingolani,
  F.~Gazeau, and T.~Pellegrino.
\newblock Thermoresponsive iron oxide nanocubes for an effective clinical
  translation of magnetic hyperthermia and heat-mediated chemotherapy.
\newblock {\em ACS Applied Materials \& Interfaces}, 2019.

\bibitem{RodriguesIJH13}
H.~F. Rodrigues, F.~M. Mello, L.~C. Branquinho, N.~Zufelato, E.~P.
  Silveira-Lacerda, and A.~F. Bakuzis.
\newblock Real-time infrared thermography detection of magnetic nanoparticle
  hyperthermia in a murine model under a non-uniform field configuration.
\newblock {\em International Journal of Hyperthermia}, 29(8):752--767, 2013.

\bibitem{BakuzisACIS13}
A.~F. Bakuzis, L.~C. Branquinho, L.~L. Castro, M.~T.~A. Eloi, and R.~Miotto.
\newblock Chain formation and aging process in biocompatible polydisperse
  ferrofluids: Experimental investigation and monte carlo simulations.
\newblock {\em Adv. Colloid Interface Sci.}, 191-192:1--21, 2013.

\bibitem{Morup1994}
S.~M\o{}rup and E.~Tronc.
\newblock Superparamagnetic relaxation of weakly interacting particles.
\newblock {\em Phys. Rev. Lett.}, 72:3278--3281, 1994.

\bibitem{BodkerPRL94}
F.~B{\o}dker, S.~M{\o}rup, and S.~Linderoth.
\newblock Surface effects in metallic iron nanoparticles.
\newblock {\em Physical Review Letters}, 72(2):282--285, 1994.

\bibitem{Dormann1988}
J~L Dormann, L~Bessais, and D~Fiorani.
\newblock A dynamic study of small interacting particles: superparamagnetic
  model and spin-glass laws.
\newblock {\em Journal of Physics C: Solid State Physics}, 21:2015--2034, 1988.

\bibitem{Dormann1996}
J~L.~Dormann, Franco D'Orazio, F~Lucari, E~Tronc, P~Prené, J~P.~Jolivet,
  D~Fiorani, R~Cherkaoui, and M~Noguès.
\newblock Thermal variation of the relaxation time of the magnetic moment of
  fe2o3 nanoparticles with interparticle interactions of various strengths.
\newblock {\em Physical review. B, Condensed matter}, 53:14291--14297, 1996.

\bibitem{BranquinhoSR13}
L.~C. Branquinho, M.~S. Carri\~{a}o, A.~S. Costa, N.~Zufelato, M.~H. Sousa,
  R.~Miotto, R.~Ivkov, and A.~F. Bakuzis.
\newblock Effect of magnetic dipolar interactions on nanoparticle heating
  efficiency: Implications for cancer hyperthermia.
\newblock {\em Scientific Reports}, 3:2887, 2013.

\bibitem{CoffeyJAP12}
W.~T. Coffey and Y.~P. Kalmykov.
\newblock Thermal fluctuations of magnetic nanoparticles: fifty years after
  brown.
\newblock {\em Journal of Applied Physics}, 112:121301, 2012.

\bibitem{knobelASP}
M.~Knobel, W.~C. Nunes, L.~M. Socolovsky, E.~De~Biasi, J.~M. Vargas, and J.~C.
  Denardin.
\newblock Superparamagnetism and other magnetic features in granular materials:
  a review on ideal and real systems.
\newblock {\em J. Nanosci. Nanotechnol.}, 8(6):2836--2857, 2008.

\bibitem{Zeb2018}
F~Zeb, M~Ishaque, K~Nadeem, M~Kamran, H~Krenn, and D~V Szabo.
\newblock Surface effects and spin glass state in co3o4 coated {MnFe}2o4
  nanoparticles.
\newblock {\em Materials Research Express}, 5:086109, 2018.

\bibitem{DenardinPRB}
J.~C. Denardin, A.~L. Brandl, M.~Knobel, P.~Panissod, A.~B. Pakhomov, H.~Liu,
  and X.~X. Zhang.
\newblock Thermoremanence and zero-field-cooled/field-cooled magnetization
  study of co(sio2)1-x granular films.
\newblock {\em Phys. Rev. B.}, 65:064422, 2002.

\bibitem{Mendon2012}
E.~Mendonça, C~B.~R.~Jesus, W.~Folly, C.~Meneses, J~Duque, and A~A.~Coelho.
\newblock Temperature dependence of coercive field of znfe2o4 nanoparticles.
\newblock {\em J. Appl. Phys.}, 111:053917, 2012.

\bibitem{BakuzisJMR96}
A.~F. Bakuzis, P.~C. Morais, and F.~A. Tourinho.
\newblock Investigation of the magnetic anisotropy in manganese ferrite
  nanoparticles using magnetic resonance.
\newblock {\em Journal of Magnetic Resonance}, 122:100--103, 1996.

\bibitem{BakuzisJAP99}
A.~F. Bakuzis, P.~C. Morais, and F.~Pelegrini.
\newblock Surface and exchange anisotropy fields in ${MnFe_2O_4}$
  nanoparticles: Size and temperature effects.
\newblock {\em Journal of Applied Physics}, 85(10):7480--7482, 1999.

\bibitem{GAZEAU1998175}
F~Gazeau, J.C Bacri, F~Gendron, R~Perzynski, Yu.L Raikher, V.I Stepanov, and
  E~Dubois.
\newblock Magnetic resonance of ferrite nanoparticles:: evidence of surface
  effects.
\newblock {\em J. Magn. Magn. Mater.}, 186(1):175 -- 187, 1998.

\bibitem{Jordan1997}
A.~Jordan, R.~Scholz, P.~Wust, H.~Fähling, J.~Krause, W.~Wlodarczyk,
  B.~Sander, Th. Vogl, and R.~Felix.
\newblock Effects of magnetic fluid hyperthermia (mfh)) on c3h mammary
  carcinoma in vivo.
\newblock {\em Int. J. Hyperthermia}, 13:587--605, 1997.

\bibitem{Rodrigues2017}
H.~F. Rodrigues, G.~Capistrano, F.~M. Mello, N.~Zufelato, E.~Silveira-Lacerda,
  and A.~F Bakuzis.
\newblock Precise determination of the heat delivery duringin vivomagnetic
  nanoparticle hyperthermia with infrared thermography.
\newblock {\em Physics in Medicine and Biology}, 62:4062--4082, 2017.

\bibitem{DennisAdvFunMat15}
C.~L. Dennis, K.~L. Krycka, J.~A. Borchers, R.~D. Desautels, J.~van Lierop,
  N.~F. Huls, A.~J. Jackson, C.~Gruettner, and R.~Ivkov.
\newblock Internal magnetic structure of nanoparticles dominates time-dependent
  relaxation processes in a magnetic field.
\newblock {\em Advanced Functional Materials}, 25(27):4300--4311, 2015.

\bibitem{Hogan2016}
S.~Jeon, K.~R~Hurley, J.~C. Bischof, C.~Haynes, and C.~Hogan.
\newblock Quantifying intra- and extracellular aggregation of iron oxide
  nanoparticles and its influence on specific absorption rate.
\newblock {\em Nanoscale}, 8:16053, 2016.

\bibitem{Bischof2014}
M.~L. Etheridge, K.~R. Hurley, J.~Zhang, S.~Jeon, H.~L. Ring, C.~J. Hogan,
  C.~L. Haynes, M.~Garwood, and J.~C. Bischof.
\newblock Accounting for biological aggregation in heating and imaging of
  magnetic nanoparticles.
\newblock {\em Tecnology}, 23:214--228, 2014.

\bibitem{Dicoroto2014}
R.~Di Corato, A.~Espinosa, L.~Lartigue, M.~Tharaud, S.~Chat, T.~Pellegrino,
  C.~Ménager, F.~Gazeau, and C.~Wilhelm.
\newblock Magnetic hyperthermia efficiency in the cellular environment
  for different nanoparticle designs.
\newblock {\em Biomaterials}, 35:6400--6411, 2014.

\bibitem{Roubeau2015}
I.~Andreu, E.~Natividad, L.~Solozábal, and O.~Roubeau.
\newblock Nano-objects for addressing the control of nanoparticle arrangement
  and performance in magnetic hyperthermia.
\newblock {\em ACS nano}, 9:1408--1419, 2015.

\bibitem{Edouard2011}
E.~Alphandéry, S.~Faure, O.~Seksek, F.~Guyot, and I.~Chebbi.
\newblock Chains of magnetosomes extracted from amb-1 magnetotactic bacteria
  for application in alternative magnetic field cancer therapy.
\newblock {\em ACS nano}, 5:6279--96, 2011.

\bibitem{Carria2017}
M.~S. Carrião, V.~R.~R.~Aquino, G.~T.~Landi, E.~L.~Verde, M.~Sousa, and A.~F.
  Bakuzis.
\newblock Giant-spin nonlinear response theory of magnetic nanoparticle
  hyperthermia: A field dependence study.
\newblock {\em Journal of Applied Physics}, 121:173901, 2017.

\bibitem{Donev2004}
A.~Donev, i.~Cisse, D.~Sachs, E.~A. Variano, F.~H. Stillinger, R.~Connelly,
  S.~Torquato, and P.M. Chaikin.
\newblock Improving the density of jammed disordered packings using ellipsoids.
\newblock {\em Science}, 303:990--993, 2004.

\bibitem{CallenJPCS}
H.~B. Callen and E.~Callen.
\newblock The present status of the temperature dependence of
  magnetocrystalline anisotropy, and the l(l+1)2 power law.
\newblock {\em J. Phys. Chem. Solids}, 27:1271--1285, 1966.

\bibitem{GleichNature05}
B.~Gleich and J.~Weizenecker.
\newblock Tomographic imaging using the nonlinear response of magnetic
  particles.
\newblock {\em Nature}, 435:1214--1217, 2005.

\bibitem{GoodwillAdvMat12}
P.~W. Goodwill, E.~U. Saritas, L.~R. Croft, T.~N. Kim, K.~M. Krishnan, D.~V.
  Schaffer, and S.~M. Conolly.
\newblock X-space mpi: Magnetic nanoparticles for safe medical imaging.
\newblock {\em Advanced Materials}, 24:3870--3877, 2012.

\bibitem{WeaverMedPhys09}
J.~B. Weaver, A.~M. Rauwerdink, and E.~W. Hansen.
\newblock Magnetic nanoparticle temperature estimation.
\newblock {\em Medical Physics}, 36(5):1822--1829, 2009.

\bibitem{ZhongSR14}
J.~Zhong, W.~Liu, L.~Kong, and P.~C. Morais.
\newblock A new approach for highly accurate, remote temperature probing using
  magnetic nanoparticles.
\newblock {\em Scientific Reports}, 4:6338, 2014.

\bibitem{NunesPRB}
W.~C. Nunes, W.~S.~D. Folly, P.~J. Sinnecker, and M.~A. Novak.
\newblock Temperature dependence of the coercive field in single-domain
  particle systems.
\newblock {\em Physical Review B.}, 70:014419, 2004.

\bibitem{GarciaJMMM}
J.~Rivas J.~Garcia-Otero, A. J. Garcia-Bastida.
\newblock Influence of temperature on the coercive field of non-interacting
  fine magnetic particles.
\newblock {\em J. Magn. Magn. Mater.}, 189:377--383, 1998.

\bibitem{Byrappa2013}
K.~Byrappa and M.~Yoshimura.
\newblock {\em Handbook of hydrothermal technology}.
\newblock Elsevier, 2001.

\bibitem{SousaMicrochem11}
M.~H. Sousa, G.~J. da~Silva, J.~Depeyrot, F.~A. Tourinho, and L.~F. Zara.
\newblock Chemical analysis of size-tailored magnetic colloids using slurry
  nebulization in icp-oes.
\newblock {\em Microchemical Journal}, 97(2):182--187, 2011.

\bibitem{MassartJMMM}
R.~Massart, V.~Dubois, E.~Cabuil, and E.~Hasmonay.
\newblock Preparation and properties of monodisperse magnetic fluids. journal
  of magnetism and magnetic materials.
\newblock {\em J. Magn. Magn. Mater.}, 149:1--5, 1995.

\bibitem{BakuzisJMMM01}
A.~F. Bakuzis and P.~C. Morais.
\newblock On the origin of the surface magnetic anisotropy in manganese-ferrite
  nanoparticles.
\newblock {\em J. Magn. Magn. Mater.}, 226-230:1924--1926, 2001.

\bibitem{EloiPRE10}
M.~T.~A. Eloi, J.~L. Santos, P.~C. Morais, and A.~F. Bakuzis.
\newblock Field-induced columnar transition of biocompatible magnetic colloids:
  An aging study by magnetotransmissivity.
\newblock {\em Physical Review E}, 82:021407, 2010.

\bibitem{KlokkenburPRL}
M.~Klokkenburg, B.~H. Ern\'e, J.~D. Meeldijk, A.~Wiedenmann, A.~V. Petukhov,
  R.~P.~A. Dullens, and A.~P. Philipse.
\newblock In situ imaging of field induced hexagonal columns in magnetite
  ferrofluids.
\newblock {\em Phys. Rev. Lett.}, 97:185702, 2006.

\bibitem{GONZAPhysRevE}
E.~S. Gon\ifmmode~\mbox{\c{c}}\else \c{c}\fi{}alves, D.~R. Cornejo, C.~L.~P.
  Oliveira, A.~M. Figueiredo~Neto, J.~Depeyrot, F.~A. Tourinho, and R.~Aquino.
\newblock Magnetic and structural study of electric double-layered ferrofluid
  with $\mathrm{MnFe_{2}O_{4}@\gamma-Fe_{2}O_{3}}$ nanoparticles of different
  mean diameters: Determination of the magnetic correlation distance.
\newblock {\em Phys. Rev. E}, 91:042317, 2015.

\end{thebibliography}

\end{document}